\documentclass[12pt,a4paper]{iopart}
\usepackage{iopams}
\usepackage{times,mathptm}
\usepackage[dvips]{graphicx}
\usepackage{enumerate}

\newcommand{\bra}{\left\langle A0\right|}
\newcommand{\ket}{\left| A0\right\rangle}

\begin{document}

\article[Migdal Theory]{TOPICAL REVIEW}{ Landau Migdal Theory of Interacting Fermi Systems:\\
A Framework for Effective Theories in Nuclear Structure Physics }
\author{Frank Gr\"ummer and Josef Speth}
\address{Institut f\"ur Kernphysik (Theorie) FZ-J\"ulich, D52425 J\"ulich, Germany}
\ead{J.Speth@fz-juelich.de}

\begin{abstract}
We review Migdal's \emph{Theory of Finite Fermi Systems} and its application to the structure of nuclei. 
The theory is an extension of Landau's \emph{Theory of Interacting Fermi Systems}. 
In the first part the basic formulas are derived within the many body Green functions approach. 
The theory is applied to isovector electric giant resonances in medium and heavy mass nuclei. 
The parameterizations of the renormalized effective ph-interaction and the effective operators are discussed. 
It is shown that the number of free parameters are restricted due to conservation laws. 
We also present an extension of Migdal's theory, where the low-lying phonons are considered in a consistent manner. 
The extended theory is again applied to the same isovector electric giant resonances and to the analysis of $(\alpha,\alpha^\prime )$ reaction data. 
We point out that the extended theory is the appropriate frame for self consistent nuclear structure calculations starting from effective Lagrangians and Hamiltonians.
\end{abstract}

\section {Introduction}
The Landau-Migdal theory of \emph{Finite Fermi Systems}\cite{Migdal67,Rev77} is the best-founded phenomenological theory for
nuclear structure calculations. It is based on Landau's \emph{Theory of Interacting Fermi Systems} \cite{LL57}. 
The basic formulas can be derived within the many-body Green functions formalism in an exact way using Landau`s renormalization procedure. 
This procedure give rise to renormalized residual interactions and effective single particle operators. 
The basic input which is required to solve the corresponding equations are the quasi particle energies, the renormalized particle-hole(ph) interaction and effective operators. 
If one considers pairing-correlations also a renormalized particle-particle interactions enters. 
So far in most of the applications the quasi particle energies have been taken from experiment and/or single-particle models and the renormalized residual interactions have been parametrized and the corresponding parameters are fitted to data. 
These \emph{Landau-Migdal parameters} turned out to be universal for all nuclei, except the lightest ones. 
More over some of these parameters are directly connected with gross properties of nuclei such as the compression modulus and the symmetry energy. 
For that reason any effective interaction, Hamiltonian or Lagrangian which is used for nuclear mean field calculations should be tested against specific Landau- Migdal parameters. 
This possibility was first used in 1975 by B\"ackman, Jackson and Speth \cite{BJS75} who showed that several parameterizations of the Skyrme forces used at that time led to an instability against spin collapse and had to be modified. 
In Ref. \cite{Krew77} this method was used to design new versions of the Skyrme interaction which could be used in self consistent nuclear structure calculations. 
In Ref. \cite{Krew88} the Landau-Migdal parameters where calculated from G-matrices based on meson exchange models. 
The authors showed in addition that the Landau expansion of the ph-interaction is well justified for all parameters except the spin-isospin parameter G' which is strongly momentum dependent due to the long range of the one-pion exchange. 
For that reason one has to include this contribution explicitly if one calculates magnetic properties \cite{BKSW80}. A detailed discussion of the Landau Migdal theory and self consistent approaches and the application of effective forces in the FFS can be found in the review by Khodel and Saperstein \cite{Kho82}.
Fully self consistent calculations in the Landau-Migdal approach have been also performed by Fayans et al. \cite{Fay00}  who based their mean field and the residual interactions on \emph{local energy-density functionals}.
Chiral Lagrangians \cite{Mei05} seems to be especially appropriate for self consistent nuclear structure calculations within the Landau-Migdal approach because they have a similar structure as the generalized Landau-Migdal interaction (which includes the one pion exchange contribution \cite{Krew88}). 
They have strong short range components and the one pion exchange is included explicitly. 
It has been shown in \cite{Krew88} that this structure of the bare interaction does not change if one performs a partial summation to obtain the  Brueckner G-matrix. 
The inclusion of the one pion exchange in chiral Lagrangians is the basic difference to Skyrme type Hamiltonians which are commonly used in self consistent nuclear structure calculations. 
This ansatz is a parametrization of a G-matrix and includes zero-range and derivatives of zero range components only. 
The zero range three body part can be replaced by a density dependent two-body interaction which simulates the density dependence of the interaction in nuclei. 
In the case of chiral Lagrangian the density dependence is introduced by the G-matrix and more over it has been shown in Ref.\cite{Mei05} that the chiral expansion also give rise to a strong three body force if applied to the nuclear many-body problem. 
It is important to mention that the single particle spectrum calculated in a  mean field approach can not be directly compared with the experimental spectrum and can therefore not be identified with Landau`s quasi particles because the coupling to phonons modifies the theoretical spectrum. 
For that reason, as we will show, the mean-field spectra are the appropriate input for the \emph{Extended Theory of Finite Fermi Systems} \cite{Rev04}, which includes in a consistent way the phonon contributions.

\section{Landau's Basic Theory}
Landau's theory of interacting Fermi systems \cite{LL57} deals with weakly excited states in infinite systems of interacting fermions, such as $^{3}$He or nuclear matter. 
Without the interaction the system would be simply a collection of independent fermions, each characterized by its momentum $\bi{p}$ and by its spin. 
The particles obey the Fermi-Dirac statistics and therefore occupy all quantum states up to the Fermi momentum $p_F$; all the states with larger momenta are unoccupied. 
The occupied states form a sphere of radius $p_F$, which is called the \emph{Fermi sphere}. 
The Fermi momentum is proportional to the number of particles

\begin{equation}\label{eq:1}
p_F=\left(\frac{3\pi^2\hbar^3N}{V}\right)^{1/3}.
\end{equation}  
Excited states are formed by moving particles from the interior to the exterior of the Fermi sphere. 
Therefore excited states differ from the ground state by a number of particle-hole pairs, the particles with $p>p_F$ and the holes with $p<p_F$. 
These particles and holes play the role of elementary excitations in the \emph{ideal Fermi gas}. 
They appear and disappear only in pairs and they have momenta close to $p_F$ for weakly excited states. 

The basic assumption by Landau is that interacting systems may be obtained from non-interacting systems by an adiabatic switching on of the interaction. 
In particular, he assumes a one-to-one correspondence between the single-particle states of the non-interacting system and the so-called quasi-particle states in the interacting system. 
These quasi-particles behave in the correlated system like real particles in the non-interacting system: They obey the Fermi-Dirac statistics and therefore occupy the corresponding quasi-particle states up to the Fermi momentum which can be calculated from the same formula (Eq.(\ref{eq:1})).
As in the non-interacting system, the excitations in the correlated system are (quasi-)particles with momenta greater than
$p_F$ and (quasi-)holes with momenta less than $p_F$ that appear and disappear, again in pairs. 
Quasi-particles are only defined near the Fermi surface, where their decay width is much smaller then their excitation energy.
In order to define such quasi-particles Landau made the following assumption concerning the interaction of the  quasi-particles: the interaction follows from a self-consistent field that acts on one specific quasi-particle and which is produced by all the other quasi-particles. 
For that reason the energy of the system is no longer the sum of the separate quasi-particle energies, but is a functional of the occupation function $n(\bi{p})$ of the quasi-particle state $\bi{p}$. 
If one excites the system one basically changes the occupation functions $n(\bi{p})$ by an amount $\delta n(\bi{p})$. 
The corresponding change of the energy is
\begin{equation}\label{eq:2} 
\delta E = \int{d\bi{p}\ \epsilon^0(\bi{p})\delta n(\bi{p})} + 
           \int{d\bi{p}\int{d\bi{p'} F^\omega(\bi{p,p'})\delta n(\bi{p})\delta n(\bi{p'})}} 
\end{equation}
where $\epsilon^0(\bi{p})$ are the equilibrium energies of the quasi-particles. 
We rewrite this expression as
\begin{equation}\label{eq:3}
\delta E = \int{d\bi{p} \left[\epsilon^0(\bi{p})+\int{d\bi{p'}F^\omega(\bi{p,p'})\delta n(\bi{p'})}\right]}\delta n(\bi{p}) = 
\int{d\bi{p}\ \epsilon(\bi{p})\delta n(\bi{p})}
\end{equation}
The natural definition of the energy of an individual quasi-particle is therefore the first variational derivative of the total energy with respect to the occupation function and that of the interaction between the quasi-particles is the second functional derivative
\begin{equation}\label{eq:4}
\epsilon(\bi{p}) = \frac{\delta E}{\delta n(\bi{p})}\qquad\qquad  
F^\omega(\bi{p,p'}) = \frac{\delta ^2 E}{\delta n(\bi{p})\delta n(\bi{p'})}.
\end{equation}
In $^{3}$He the interaction $F^\omega$ has a spin-independent and a spin-dependent part. 
\begin{equation}\label{eq:5}
F^\omega(\bi{p,p'}) = F(\bi{p,p'}) + G(\bi{p,p'}) \bsigma\cdot\bsigma'
\end{equation}
The energies of the elementary excitations are measured from the Fermi sphere. 
For quasi-particles one obtains
\begin{equation}\label{eq:6}
\xi(p) = \frac{p^2}{2m}-\frac{p^2_F}{2m}\approx v(p-p_F)
\end{equation}
and for quasi-holes
\begin{equation}\label{eq:7}
-\xi(p) = \frac{p^2_F}{2m}-\frac{p^2}{2m} \approx v(p_F-p)
\end{equation}
The velocity of the excitations at the Fermi surface is given by 
\begin{equation}\label{eq:8}
v=\frac{p_F}{m^*}
\end{equation}
As we shall see later the effective mass $m^*$ is directly connected with $F^\omega$.

If one considers excitations near the Fermi surface one may restrict the magnitude of the momenta $\bi{p}$ and $\bi{p'}$ in the interaction $F^\omega(\bi{p,p'})$ to the Fermi momentum. 
The only variable then is the angle $\Theta$ between the two momenta. 
This suggests an expansion in Legendre polynomials
\begin{equation}\label{eq:9}
f(\bi{p,p'}) = \sum_{l}{f_{l}\ P_{l}(\cos\Theta)}
\end{equation}
The constants $f_l$ are the famous Landau parameters. 
One introduces dimensionless parameters by defining
\begin{equation}\label{eq:10}
F_{l} = C_{0}f_{l}
\end{equation}
where
\begin{equation}\label{eq:11}
C_{0} = \left(\frac{dn}{d\epsilon}\right)^{-1} 
\end{equation} 
is the inverse density of states at the Fermi surface. 
Likewise, one may also expand the other component of the interaction in eq.(\ref{eq:5}) and define the parameter $g_{l}$ in the same way.
The parameters $f_{0}$ and $f_{1}$ are related to the compression modulus $K$ by
\begin{equation}\label{eq:12}
K = p^2_{F}\frac{d^{2}(E/A)}{dp^{2}_{F}} = 6\frac{\hbar^2p ^2_F}{2m^*}\left(1+2f_0\right) 
\end{equation}
and the effective mass $m^*$
\begin{equation}\label{eq:13}
m^*/m = 1 + \frac{2}{3}f_{1}. 
\end{equation}
The interacting quasi-particles and quasi-holes in $^3$He give rise to the famous zero sound and spin waves. 
The corresponding linearized time dependent equation for $\delta n(\bi{p})$ is equivalent to the random phase approximation (RPA) equation; it is the basis of the Landau- Migdal theory.

\section{Migdal's Theory of Finite Fermi Systems}
Migdal has applied Landau's \emph{theory of interacting Fermi system} to atomic nuclei \cite{Migdal67}. 
Here one deals with two kinds of fermions --\emph{protons and neutrons}-- and, more importantly, with a relatively small number of particles.
In the past 40 years Migdal's theory has been successfully applied to numerous nuclear structure problems. 
Whereas in infinite systems the quasi-particles differ from the bare particles only by their effective mass, the situation in finite nuclei is completely different. 
Here the quasi-particles are the single particle excitations of the nuclei, which can be obtained from phenomenological shell models or, in the case of self consistent approaches, by the corresponding mean-field solutions of a given effective Lagrangian, effective Hamiltonian and energy density functionals, respectively.
The quasi-particle interaction is defined in the same way as before.
The interaction, however, depends not only on the spin but also on the isospin of the quasi-particles and is density dependent because the interaction inside a nucleus is very different from the interaction in free space. 
As in infinite systems, one expands the interaction at the Fermi surface in terms of Legendre polynomials and the parameters of the expansion --the famous \emph{Landau-Migdal parameters}-- are determined from experiment. 
These parameters turn out to be universal for all nuclei but the very light ones.

As pointed out in the introduction,the \emph{Theory of Finite Fermi Systems (TFFS)} is a semi-phenomenological microscopic theory. 
It is microscopic because all its equations are derived rigorously from first principles, however it contains also phenomenological aspects, such as the quasi-particle energies and wave functions and the quasi-particle--quasi-hole interaction. All these quantities are defined microscopically and could, in principle, be calculated starting from the bare nucleon-nucleon interaction. 
Such calculations are very involved and in actual calculations one has to make severe approximations, so that we cannot expect to obtain quantitative agreement between, for example, the calculated and phenomenological Landau-Migdal parameters. 
Nevertheless the calculated Landau-Migdal parameters are in surprisingly good agreement with the phenomenological ones \cite{Krew88}.
In the TFFS, as in Landau's theory, one derives a renormalized equation for the response functions and solves this equation in a given approximation.
For that reason the TFFS can be applied to such problems as are connected with the response function.
These include, for example, excitation energies and transition probabilities in even nuclei and moments and transitions in odd-mass nuclei.
The input data. single particle energies and wave functions are supposed to be given by a shell model in the phenomenological approach, or are calculated in mean field approximation in self consistent calculations. Similarly the residual particle-hole and particle -particle interactions are either parametrized and adjusted to the experiment or determined from the corresponding effective Lagrangian and effective Hamiltonian, respectively.
In this respect is the TFFS and its extension (which includes phonons explicitly) a frame for any nuclear structure calculations, for self consistent and non self consistent ones.
This will be discussed in detail in the following chapters.

\subsection{Basic properties of many body Green functions}
In this section we will give a short introduction to the theory of many-body Green functions. (More details are given in \cite{Rev77}). 
They are the basis of the TFFS.
The starting point of the theory is a Hamiltonian for Fermions
\begin{equation}\label{eq:14} 
H = \sum_\nu \epsilon_\nu^0\; a_\nu^+a_\nu + \frac{1}{4} \sum_{\nu_1 ... \nu_4} w{{_{\nu_2}}{_{\nu_4}} {_{\nu_1}}{_{\nu_3}}}\;a_{\nu_4}^+a_{\nu_2}^+ a_{\nu_1}a_{\nu_3}
\end{equation}
Here $\epsilon_\nu^0$ are  single particle energies, while $a_{\nu}^+$ and $a_\nu$ are creation and annihilation operators of nucleons in the state $\varphi_\nu $, which obey anticommutation rules, and \emph{w} is the residual two-body interaction.
We now consider an $A$-particle system
\begin{equation}\label{eq:15}
H\left|An \right\rangle =E{^A}\!\!{_n}\left|An \right\rangle
\end{equation}
where $\left|An \right\rangle$ and $E{^A}\!\!{_n}$ are assumed to be the exact solutions of the many-body Schr\"odinger equation with a given many-body Hamiltonian.
The one-particle and two-particle Green functions are defined as
\begin{equation}\label{eq:16}
g_{\nu_1\nu_2}\;{(t_1;t_2)} = (-i)\bra{T{\left\{{a_{\nu_1}{(t_1)}{a_{\nu_2}^+{(t_2)}}}\right\}}}\ket
\end{equation}
\begin{equation}\label{eq:17}
g_{\nu_1\nu_3\nu_2\nu_4}\;{(t_1t_3;t_2t_4)} =
(-i)^2)\bra{T{\left\{{a_{\nu_1}{(t_1)}{a_{\nu_3}{(t_3)}}{a_{\nu_4}^+{(t_4)}}{a_{\nu_2}^+{(t_2)}}}\right\}}}\ket
\end{equation}
with the time-dependent creation and annihilation operators
\begin{equation}\label{eq:18}
a_{\nu}(t)= e^{iHt}\;a_{\nu}\;e^{-iHt}\qquad\qquad a_{\nu}^+(t)= e^{iHt}\;a_{\nu}^+\;e^{-iHt}
\end{equation}
The symbol $T$ denotes the time-ordering operator of Wick, which means that the operators should be taken in time-ordered form with the latest time to the left and the earliest to the right. 
In addition, the product has to be multiplied by a factor $\pm 1$ depending on whether an even or odd number of permutations is needed to come from the given to the time-ordered form.
The Fourier transform of the one-particle Green function has the form
\begin{equation}\label{eq:19}
g_{\nu_1\nu_2}\;{(t_1;t_2)}\;=
\frac{1}{2\pi}\int_{-\infty}^{+\infty}\;d\omega{\;}{g_{\nu_1\nu_2}}{(\omega)}\;e^{-i\omega{(t_1-t_2)}}
\end{equation}
Correspondingly, the two-particle Green function depends on three frequencies.
For further investigation it is convenient to introduce Green functions with a source field $q$
\begin{equation}\label{eq:20}
g_q{(1,2)}\;= 
\frac{(-i)\bra{T{\left\{U{a_{\nu_1}{(t_1)}{a_{\nu_2}^+{(t_2)}}}\right\}}}\ket}{\bra{T{\left\{U\right\}}}\ket}
\end{equation}
with 
\begin{equation}\label{eq:21}
U=e^{i\int{d5d6\;q(5,6){a_{\nu_5}^+{(t_5)}{a_{\nu_6}{(t_6)}}}}}
\end{equation}
and the short hand notation:
\begin{equation}\label{eq:21a}
\sum_{\nu_1}\int{dt_1} = \int{d1}.
\end{equation}

In order to determine the Green functions one needs the equations of motion. 
They can be obtained by differentiation of the corresponding Green function with respect to time. 
For the one-particle Green function with a source field we obtain
\begin{equation}\label{eq:22}\fl
\int{d3 \left\{S(1,3)\;+\;q(1,3)\right\}g_q{(3,2)}}+
\frac{i}{2}\int{d3d4d5\;w(14,35)\;g_q{(35,24)}} = \delta(1,2)
\end{equation}
with the abbreviations
\begin{eqnarray}\label{eq:23}
\delta_{\nu_1\nu_2}\delta(t_1-t_2)\left\{i\frac{\delta}{\delta{t_1}}-\epsilon_{\nu_1}^0\right\} &= S(1,2).
\end{eqnarray}
Here we have used eq.(\ref{eq:14}) in order to eliminate the Hamiltonian $H$ in eq.(\ref{eq:22}). 
The equation for the two-body Green function is connected with the three-body Green function and so forth. 
Thereby one obtains a hierarchy of equations. 
In order to calculate the one-body Green functions one needs the two-body Green function, etc. 
This is not the way that we proceed in the following; rather, we replace the bare two-body interaction $w$ by an effective one-body potential $\Sigma$ (\textit{mass operator or self-energy}) which, in principle, is given by the bare interaction and the two-particle Green function. 
The corresponding equation for the one-particle Green function is called the Dyson equation.
\begin{equation}\label{eq:24}
\frac{i}{2}\int{d3\left\{S(1,3)\;+\;q(1,3)\;+\;\Sigma_q{(1,3)}\right\}g_q{(3,2)}}=\delta (1,2)
\end{equation}
So far we have only replaced the unknown two-particle Green function by the also unknown self-energy $\Sigma$ and little seems to be gained. 
Before we discuss the solution of this problem we continue with the development of the formal theory.
We define the response function $L$ by the functional derivative of the one-particle Green function
\begin{equation}\label{eq:25}
\frac{\delta g_q{(1,2)}}{\delta q(4,3)}\;=\;L(13,24)\;=\;g(13,24)\;-\;g(1,2)g(3,4) 
\end{equation}
The functional derivative of the Dyson equation gives an integral equation for the response function
\begin{equation}\label{eq:26}\fl
L(13,24)= -g(1,4)g(3,2) -i\int{d5d6d7d8\;g(1,5)K(57,68)L(83,74)g(6,2)}
\end{equation}
where we introduced the effective two-body interaction $K$ via
\begin{equation}\label{eq:27}
\frac{\delta \Sigma (1,2)}{\delta g(3,4)}=iK(13,24)
\end{equation}
\begin{figure}[htbp]
\begin{center}
\includegraphics[bb=45 504 606 560,width=12cm]{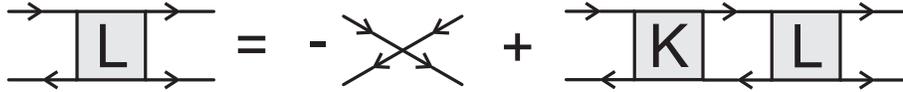}
\end{center}
\caption{\label{fig:1}
Graphical representation of the Bethe-Salpeter equation for the response function in the ph channel.}
\end{figure}
A graphical representation of eq.(\ref{eq:26}) is given in Fig.\ref{fig:1}. 
One notices that the kernel $K$ of the integral equation is irreducible with respect to the particle-hole propagator.
We further introduce the vertex function $\widetilde{\tau}$ by
\begin{equation}\label{eq:28}
L(13,24)= \int{d5d6\;g(1,5)\;\widetilde{\tau}{(53,64)}\; g(6,2)}
\end{equation}
The corresponding equation follows from eq.(\ref{eq:26})
\begin{equation}\label{eq:29}\fl
{\widetilde{\tau}{(13,24)}}=-\delta (1,4)\delta (3,2)
-i\int{d5d6d7d8\;K(17,28)g(8,5)g(6,7)\widetilde{\tau}(53,64)}
\end{equation}
The linear response function allows one to determine the following quantities :
\begin{enumerate}[(I)]
\item
From eq.(\ref{eq:25}) we obtain in linear approximation the change of the one-particle Green function :
\begin{equation}\label{eq:30}
\delta g_q{(1,2)}\;=\;\sum_{3,4}L(13,24)\delta q(4,3)
\end{equation}
The Fourier transformation of the time variables gives $\delta g_q{(\omega ,\Omega)}$ and, after integration over $\omega$, one obtains $\delta \widetilde{\rho}_q (\Omega)$, where $\Omega$ is the energy transferred by the external field $\delta q (\Omega)$ and $\delta \widetilde{\rho}_q (\Omega)$ is the change of the density due to this external field in a given single particle basis $\varphi_{\nu}$ :
\begin{equation}\label{eq:31}
\delta \widetilde{\rho}_{\nu_1 \nu_2}{(\Omega)}\;=\;\sum_{\nu_3,\nu_4} L(\Omega)_{\nu_1 \nu_3 \nu_2 \nu_4}\delta q_{\nu_4 \nu_3}(\Omega)
\end{equation}
\item
With the time order $t_3,t_4 > t_1,t_2$, we insert into the two particle Green function a complete system of eigenfunctions of the A-particle system between the two particle-hole pairs
\begin{equation}\label{eq:32}
g(13,24)=\sum_{n=0}^\infty\;g^{0n}(34)g^{n0}(12)
\end{equation}
with
\begin{equation}\label{eq:33}
g^{0n}(34)\;=\;-i \left\langle A0 \right| T{\left\{ {a(3)a^+(4)}\right\}} {\left| An\right\rangle}
\end{equation}
and
\begin{equation}\label{eq:34}
g^{n0}(12)\;=\;-i \left\langle An \right| T{\left\{ {a(1)a^+(2)}\right\}} {\left| A0\right\rangle}
\end{equation}
As the response function is defined as
\begin{equation}\label{eq:35}
L(13,24)\;=g(13,24)\;-g(34)g(12)
\end{equation}
its spectral representation is also given by eq.(\ref{eq:32}) but the sum begins at $n=1$.
With the appropriate time order one obtains from the generalized Green function in eq.(\ref{eq:34}) the transition amplitudes between the ground state and the excited states of the A-particle system.
\item
With the time order $t_3 \rightarrow \pm \infty$ and $t_4 \rightarrow \mp\infty$ in the two particle Green function one can move these operators to left and right side of the time ordering and may introduce two complete sets of eigenstates of the (A+1)- system or (A-1)-system, respectively. 
From this one obtains another generalized Green function
\begin{equation}\label{eq:36}
g^{\alpha,\beta}(12)\;=\;-i \left\langle A\pm1,\alpha \right| T{\left\{ {a(1)a^+(2)}\right\}} {\left| A\pm1,\beta\right\rangle}.
\end{equation}
where $\alpha$ and $\beta$ are two given states in the odd mass system.
As we know only the response function, one has to subtract the one-particle Green function of the A-particle system
and obtains finally the change of the density between two given states in the ($A\pm1$)-system and the ground state of the A-particle system. 
With this quantity one can calculate moments and transitions in odd-mass nuclei. Details are given in \cite{Rev77}.
\end{enumerate}

\subsection{One-particle Green function and the nuclear shell model}
In order to solve the equation for the response function and vertex function, respectively, we have to know the one-particle Green functions $g(1,2)$ and the effective two-body interaction $K$.
As Landau has shown, one does not need the full information included in $g$ but only the so-called $\emph pole\; part$ of the one-body Green functions near the Fermi surface. 
The rest gives rise to a renormalized interaction and effective one-body operators.
We first discuss the pole part of the one-body Green function. 
For that we consider the Dyson equation in coordinate space \cite{Migdal67}
\begin{equation}\label{eq:37}
\left({\epsilon -\frac{p^2}{2m}}\right)g{\left({\xi_1,\xi_2,\epsilon}\right)}
-\int{d \xi_3\;{\Sigma(\xi_1,\xi_3,\epsilon)}}\;g(\xi_3,\xi_2,\epsilon)=\delta ({\xi_1}-\xi_2)
\end{equation}
where $\xi\equiv{(\bi{r},s)}$ represents space and spin coordinates.
In an arbitrary single-particle basis $\widetilde {\varphi}_{\nu}(\xi)$ the Dyson equation has the form 
\begin{equation}\label{eq:38}
\epsilon \;g_{\nu_1\nu_2}-\sum_{\nu_3}\left[\frac{p^2}{2m} +\Sigma(\xi_1,\xi_3,\epsilon)\right]_{\nu_1\nu_3}g_{\nu_3\nu_2}=\delta_{\nu_1\nu_2}
\end{equation}
We now chose a special basis $\varphi_{\nu}(\xi)$ that diagonalizes the expression in the brackets
\begin{equation}\label{eq:39}
\left[\frac{p^2}{2m} +\Sigma(\xi_1,\xi_3,\epsilon)\right]_{\nu_1\nu_3}= E_{\nu_1}(\epsilon)\delta_{\nu_1\nu_3}
\end{equation}
It is obvious that such a basis has to depend on the energy $\epsilon$.
The one-body Green function becomes diagonal in this basis and can be written as
\begin{equation}\label{eq:40}
g_{{\nu_1}\nu_2}=\frac{{\delta}_{\nu _1 \nu_2}}{\epsilon - E_{\nu_1 }(\epsilon)}
\end{equation}
The functions $\varphi_\nu$ that diagonalize $g$ satisfy the equation
\begin{equation}\label{eq:41}
\left[\frac{p^2}{2m} +\Sigma(\xi,\bi{p},\epsilon)\right]\varphi _{\nu}(\xi,\epsilon)
=E_{\nu}(\epsilon)\varphi_{\nu}(\xi,\epsilon)
\end{equation}
Here the self-energy $\Sigma$ is given in a "mixed" representation, where we have replaced the non-locality by a 
momentum dependence
\begin{equation}\label{eq:42}
\Sigma {(\xi ,\bi{p},\epsilon)}=\int\ d\xi_1 \Sigma {(\xi, \xi{_1},\epsilon)} \;e^{i(\bi{\xi}-\bi{\xi}_{1})\bi{p}}
\end{equation}
Due to Landau's renormalization procedure one needs the one-particle Green functions only near the poles $\epsilon_\nu$, with
\begin{equation}\label{eq:43}
\epsilon_\nu = 
E_{\nu}(\epsilon_{\nu)}=\left[\frac{p^2}{2m} +\Sigma(\xi,\bi{p},\epsilon_\nu)\right] _{\nu \nu}
\end{equation}
The single-particle energies $\epsilon_\nu$ so defined are the quasi-particle energies (in the sense of Landau) for finite systems.
We expand eq.(\ref{eq:41}) around the quasi-particle energy $\epsilon_\nu$ up to first order in $\epsilon$ and obtain for the one-(quasi-)particle Green function
\begin{equation}\label{eq:44}
g_{\nu_1 \nu_2}(\epsilon) 
= \frac{ \delta _{\nu _1 \nu_2}}{ \epsilon - \epsilon_{\nu_1} +i\gamma_{\nu_1}}\;
 {\frac{1}{\left(1-{\frac{dE_\lambda}{d\epsilon}}\right)}}_{\epsilon=\epsilon_{\nu_1}}
\end{equation}
The residue of $g$ at the pole is called the \textit{single particle strength}
\begin{equation}\label{eq:45}
z_\nu ={\frac{1}{\left(1-{\frac{dE_\nu}{d\epsilon}}\right)}}_{\epsilon=\epsilon_{\nu_1}}
\end{equation}
As we expanded the energy up to first order around the pole, one needs the wave functions only at the pole energy,
$\varphi_{\nu} (\xi) = \varphi_{\nu} (\xi,\epsilon_\nu)$.
These wave functions satisfy the equation
\begin{equation}\label{eq:46}
\left[\frac{p^2}{2m} +\Sigma(\xi,\bi{p},\epsilon_\nu)\right] \varphi _{\nu}(\xi)=
\epsilon _{\nu} \varphi _{\nu}(\xi)
\end{equation}
We introduce the quantity $p^{2}_F$, which is defined by
\begin{equation}\label{eq:47}
\frac{p^{2}_F (\bi{r})}{2m} + \Sigma(\bi{r},p_{F}(\bi{r}),\epsilon_F) = \epsilon_F
\end{equation}
where $\epsilon_F$ is the Fermi energy. 
In infinite Fermi systems $p_F$ can be identified with the momentum at the Fermi surface.
We now expand the self-energy $\Sigma $ about the Fermi momentum $p_{F}$ and the Fermi energy $\epsilon_F$
\begin{equation}\label{eq:48}
\left\{ \left( \frac{1}{2m} +\frac{\delta \Sigma}{\delta p^2}\right)_F \left[ p^2 -p^2{_F} (\bi r)\right] -\left(1-\frac{\delta \Sigma}{\delta \epsilon}\right)_F ( \epsilon _{\lambda } -\epsilon_F )\right\}\varphi_{\lambda}=0
\end{equation}
This formula can be rewritten as
\begin{equation}\label{eq:49}
\left\{\frac{1}{2m^*} \; p^2 + U(\textbf r)\right\} \varphi _\nu = 
\epsilon_{\nu}\varphi _\nu
\end{equation}
with the abbreviations
\begin{equation}\label{eq:50}
\frac{m}{m^*} =
\frac{\left(1+2m \frac{\delta \Sigma}{\delta p^2}\right)_{F}} 
{\left({1-\frac{\delta \Sigma}{\delta \epsilon}}\right)_F}
\end{equation}
and 
\begin{equation}\label{eq:51}
U( \bi{r}) = 
\mu -\frac{p^2{_F} ( \bi{r})}{2m^*} = 
\frac{m}{m^*}\;\left( \Sigma\right)_F + \mu \left( 1-\frac{m}{m^*}\right)
\end{equation}
If we also take the spin degree of freedom into account then, up to order $\bi{p}^2 $ in the self-energy $\Sigma$, one obtains the following single-particle Hamiltonian
\begin{equation}\label{eq:52}
H=\frac{1}{2m^*} \; p^2 + U(\bi{r}) + \alpha(\bi{r} ) \bsigma\cdot\bi{l} + \beta(\bi{r}) l(l+1)
\end{equation}
Here $\textbf l$ is the angular momentum operator and $l$ its quantum numbers.
The form of this Hamiltonian follows from the expansion of the self-energy $\Sigma$ about the Fermi energy $\epsilon_F$ and the Fermi momentum $p_F$, and symmetry arguments.
This Hamiltonian agrees with phenomenological shell model Hamiltonians.
It is useful to point out that the spin-orbit term follows here from the expansion of the non-locality of $\Sigma$.

\subsection{Landau's renormalization procedure}
We now return to the one-particle Green function.
With eqs. (\ref{eq:44} and \ref{eq:45}) we can separate the one-particle Green function into a singular part and a remainder
\begin{equation}\label{eq:53}
g_{\nu_{1}\nu_{2}}(\omega)=\frac{\delta_{\nu_{1}\nu_{2}}z_{\nu_1}}{\omega - \epsilon_{\nu_1} +i \eta \; sign ( \epsilon_{\nu_1}- \mu)} \; +\;g^{(r)}_{\nu_{1}\nu_{2}}(\omega)
\end{equation}
with
\begin{eqnarray}\label{eq:54}\nonumber
z_{\nu_1}={\left|\left\langle A0\left|a_{\nu_1}\right|A\pm1\;\nu{_1}\ \right\rangle\right|}^2 \\
\epsilon_{\nu_1}= E^{A+1}_{\nu _1}-E^{A}_0  \;\;\;\; for \;\;\epsilon_{\nu_1}> \mu \\ \nonumber
\epsilon_{\nu_1}= E^{A}_0-E^{A-1}_{\nu _1}  \;\;\;\; for \;\;\epsilon_{\nu_1}< \mu
\end{eqnarray}
With this \textit{ansatz} one writes the product of two one-particle Green functions --a quantity that enters in all the integral equations we have derived before-- as a singular part $S$ and the remainder $R$, using the \textit{ansatz} in eq.(\ref{eq:53})
\begin{equation}\label{eq:55}
g_{\nu_1 \nu_3}\left(\omega +\frac{1}{2}\Omega\right)  g_{\nu_2\nu_4}\left(\omega -\frac{1}{2}\Omega\right)=
S_{\nu_1\nu_2 \nu_3 \nu_4}\left( \omega; \Omega\right) +  R_{\nu_1 \nu_2 \nu_3 \nu_4}\left( \omega; \Omega\right)       \end{equation}
with
\begin{equation}\label{eq:56}
S_{\nu_1 \nu_2 \nu_3 \nu_4}\left( \omega; \Omega\right)=
2 \pi i z_ {\nu_3} z_{ \nu_4} \delta_{ \nu_1 \nu_3} \delta_{ \nu_2 \nu_4} \; \delta \left( \omega - \frac{\epsilon_{ \nu_3} + \epsilon_{ \nu_4}} {2}\right)\;\frac{n_{\nu_2}-n_{\nu_1}}{\epsilon_{\nu_2}- \epsilon_{\nu_1}-\Omega}
\end{equation}
Using Landau's renormalization procedure one obtains an equation for the renormalized vertex function $\tau$
\begin{eqnarray}\label{eq:57}
\tau_{\nu_1\nu_3, \nu_2 \nu_4 }\left( \omega ,\Omega\right)=\tau^{\omega}_{\nu_1 \nu_3, \nu_2 \nu_4 }\left( \omega , \Omega\right) +\\ \nonumber
\sum_{\nu_5 \nu_6}F^{ph}_{\nu_1 \nu_5 , \nu_2 \nu_6}\left( \omega,\frac{\epsilon_{\nu_5}+ \epsilon_ {\nu_6}}{2}, \Omega\right) \; \frac{n_{\nu_5} -n_{\nu_6}}{\epsilon_{\nu_5} - \epsilon_{ \nu_6} -\Omega } \;\tau_{\nu_5 \nu_3, \nu_6 \nu_4 }\left( \frac{\epsilon_ {\nu_5} + \epsilon_{\nu_6}}{2} ;\Omega\right)
\end{eqnarray}
The renormalized vertex is defined as:
\begin{equation}
\tau_{\nu_1\nu_3, \nu_2 \nu_4 }\left( \omega ,\Omega\right)=\sqrt{z _{\nu _1} z_{\nu _2}} \; \widetilde{\tau }_{\nu_1\nu_3, \nu_2 \nu_4 }\left( \omega ,\Omega\right)
\end{equation}
Here only the singular part of the product of the two Green functions appears explicitly, whereas the remainder gives rise to a renormalized effective two-body interaction $F^{ph}$ and a renormalized inhomogeneous term $\tau^\omega$. 
In similar fashion one can renormalize the response function $L$ and obtains
\begin{equation}\label{eq:58}
L(13,24)=\tau ^{\omega} (13,57) \widetilde{L}(57,68)\tau^{\omega}(68,24) +\tau ^{\omega} (13,57)R(57,24)
\end{equation}
where $\widetilde{L}$ is the quasi-particle response function
\begin{equation}\label{eq:59}
\widetilde{L}(13,24) = S(1,2) \tau(1324)
\end{equation}

\subsection{The renormalized equations of the TFFS}
As we have seen in section 2.1, the response function includes the transition amplitudes between the ground state and the excited states of an A-particle system. 
Using the arguments in section 2.1 and the projection procedure described in \cite{Rev77}, one obtains from the renormalized response function in eq.(\ref{eq:58}) and the renormalized vertex function eq.(\ref{eq:57}) an equation for the quasi-particle quasi-hole transition matrix elements
\begin{equation}\label{eq:60}
\left( \epsilon_{\nu_1}-\epsilon_{\nu_2}- \Omega\right)\chi^{m}_{\nu_1 \nu_2} =
\left(n_{\nu_1}-n_{\nu_2}\right)\sum_{\nu_3 \nu_4}F^{ph}_{\nu_1 \nu_4 \nu_2 \nu_3} \;\chi^{m}_{\nu_3 \nu_4}
\end{equation}
which are connected with the full ph-transition matrix elements
\begin{eqnarray}\label{eq:61}
\chi^{m0}_{\nu_1 \nu_2} = < Am|a^{+}_{\nu_1} a_{ \nu_2}|A0>
\end{eqnarray}
by the relation 
\begin{equation}\label{eq:62}
\chi^{m0}_{\nu_1 \nu_2}= \sum_{\nu_3 \nu_4}\tau^{\omega}_{\nu_3 \nu_1 \nu_4 \nu_2} \;\chi^{m}_{\nu_3 \nu_4}
\end{equation}
The transition matrix element of a one-body operator $Q$ is given by
\begin{equation}\label{eq:63}
<Am|Q|A0> \;=\sum_{\nu_1 \nu_2}Q^{\rm eff}_{\nu_1 \nu_2}\; \chi ^{m}_{\nu_1 \nu_2}
\end{equation} 
where $Q^{\rm eff}$ is the renormalized one-particle operator
\begin{equation}\label{eq:64}
Q^{\rm eff}_{\nu_1 \nu_2}=\sum_{\nu_3 \nu_4}\tau^{\omega}_{\nu_1 \nu_3 \nu_2\nu_4}Q_{ \nu_3 \nu_4}
\end{equation}
In section 2.1 we also showed that the response function allows one to calculate the change of the density in an external field.
\begin{equation}\label{eq:65}
\delta \widetilde{\rho}_{\nu_1 \nu_2}{(\Omega)}\;=\;\sum_{\nu_3,\nu_4} L_{\nu_1 \nu_2 \nu_3 \nu_4}(\Omega)\;\delta q_{\nu_3 \nu_4}(\Omega)
\end{equation}
In the following we are interested in the transition strength as a function of the excitation energy. 
The expression for the strength function is given by
\begin{equation}\label{eq:66}
S(\Omega ,\Delta)= -\frac{i}{\pi}Im\sum_{\nu_1 \nu_2} Q_{\nu_1 \nu_2}(\Omega)\; \delta \widetilde{\rho} _{\nu_2 \nu_1}(\Omega+i\Delta)
\end{equation}
where $\Delta $ is a smearing parameter that simulates the finite experimental resolution and the more complex configurations not included in the present approach.
From eqs.(\ref{eq:57}) to (\ref{eq:59}) and (\ref{eq:64})
we obtain, with $\delta q \equiv Q$ and $\delta \widetilde{\rho} \equiv \widetilde {\rho}$,
\begin{equation}\label{eq:67}
S(\Omega ,\Delta)= -\frac{i}{\pi}Im\sum_{\nu_1 \nu_2} Q^{\rm eff}_{\nu_1 \nu_2}(\Omega) {\rho} _{\nu_2 \nu_1}(\Omega +i\Delta)
\end{equation}
where $\rho$ is the change of the quasi-particle density
\begin{equation}\label{eq:68}
{\rho}_{\nu_1 \nu_2}(\Omega)=  \frac{n_{\nu_1} -n_{\nu_2}}{\epsilon_{\nu_1} - \epsilon_ {\nu_2} -\Omega }Q^{\rm eff}_{\nu_1 \nu_2}(\Omega)- \; \frac{n_{\nu_1} -n_{\nu_2}}{\epsilon_{\nu_1} - \epsilon_ {\nu_2} -\Omega }\sum_{\nu_3 \nu_4}F^{ph}_{\nu_1 \nu_3 \nu_2 \nu_4} \rho_{\nu_4 \nu_3}(\Omega)
\end{equation}
The second term on the right side of eq.(\ref{eq:58}) can be neglected because we are interested in the singular solutions (resonances) of the equation. The calculations presented here have been performed in $\bf{r}$-space because this allows treatment of the continuum in the most efficient way. The corresponding equations have the form \cite{Tsel93,Rev04}
\begin{eqnarray}\label{eq:69}
\rho(\bi{r},\Omega) = &-\int d^{3}\bi{r}' A(\bi{r},\bi{r}',\Omega) Q^{\rm eff}(\bi{r}',\Omega) \\ \nonumber
&-\int d^3\bi{r}'d^3\bi{r}''A(\bi{r},\bi{r}',\Omega) F^{ph}(\bi{r}',\bi{r}'') \rho(\bi{r}'',\Omega)
\end{eqnarray}
with the ph-propagator 
\begin{equation}\label{eq:70}
A(\bi{r},\bi{r}',\Omega)=\sum_{{\nu_1},{\nu_2}}\frac{n_{\nu_1} -n_{\nu_2}}{\epsilon_{\nu_1} - \epsilon_{\nu_2} -\Omega }\phi^{*}_{\nu_1}(\bi{r})\phi^{*}_{\nu_2}(\bi{r}')\phi_{\nu_1}(\bi{r})\phi_{\nu_2}(\bi{r}')
\end{equation}
and the distribution of the transition strength
\begin{equation}\label{eq:71}
S(\Omega, \Delta) = -\frac{1}{\pi}Im \int d^{3}\bi{r} Q^{\rm eff}{(\bi{r},\Omega)}^{*} \rho(\bi{r},\Omega +i\Delta )
\end{equation}
These formulas refer to even-even nuclei because only in that case can one define the Landau quasi-particles as the single-particle excitations in the neighboring odd mass nuclei.
This holds for spherical as well as for deformed nuclei \cite{ Sp70}.

We also have shown that the theory allows one to calculate moments and transitions in odd mass $ (A\pm1)$ nuclei, where A refers in general to a closed-shell nucleus.
We then obtain for example, 
\begin{equation}\label{eq:72}
<A\pm1, \alpha|Q|\beta,A\pm1>-<A,0|Q|0,A> \delta_{\alpha \beta}=
\sum_{\nu_1 \nu_2}Q^{\rm eff}_{\nu_1 \nu_2} \delta {\rho}^{\alpha \beta}_{\nu_1 \nu_2}
\end{equation}
with
\begin{equation}\label{eq:73}
{\rho}^{\alpha \beta}_{\nu_1 \nu_2}=\delta_{\nu_1 \alpha} \delta_{\nu_2 \beta}+
\frac{n_{\nu_1} -n_{\nu_2}}{\epsilon_{\nu_1} - \epsilon_ {\nu_2} -\Omega_{\alpha \beta} } \sum_{\nu_3 \nu_4}F^{ph}_{\nu_{1} \nu_{3} , \nu_{2} \nu_4} {\rho}^{\alpha \beta}_{\nu_3 \nu_4}
\end{equation}
The derivation is given in \cite{Rev77}. Eqs. (\ref {eq:67}) and (\ref{eq:72}) are the basis of the TFFS.

\subsection{Application to the giant electric dipole resonances in nuclei}
As examples for the TFFS and --later-- for the ETFFS, we present numerical results for the giant electric dipole resonance in the doubly magic nuclei $^{40}$Ca,$^{132}$Sn and $^{208}$Pb. 
The calculations presented here are performed in the $\bi{r}$ -space and not in the configuration space of a shell model. 
In the $\bi{r}$ -space one is able to include all single particle states which can contribute to a given multipolarity if the TFFS is applied, for example, to collective states in nuclei. 
The single particle continuum is included correctly using the method suggested by Shlomo and Bertsch \cite{Shlo75}. 
We wish to stress that the TFFS equation and the continuum RPA equation are formally identical. 
The TFFS equations are derived in a rigorous manner, therefore they are \emph{exact} until one choses certain approximations for the quasi-particles, the interaction and the effective operators. 
As we have exact relations we are able to apply conservation laws, which connect the interaction parameters with gross properties of nuclei (symmetry energy, compression modulus and effective mass) and which give information about the renormalized operators (Ward identities). In addition, the Green function method is the most appropriate way to extend the theory in a consistent manner, as we will show in the next section. In the present calculations the experimental single particle energies are used as far as possible, and for the rest we take the solutions of a Woods-Saxon shell model potential, the parameters of which are adjusted to reproduce the experimentally known energies, as discussed in \cite{Rev04}.

For the calculation of the electric states we used the conventional Landau-Migdal parametrization for the ph-interaction:
\begin{equation}\label{eq:74}\fl
F^{ph}\left(1,2\right) = C_0 \delta\left(\bi{r}_1 - \bi{r}_2\right)[f_0 (\rho)+f'_0 (\rho)\btau\cdot\btau'
+g_0 (\rho)\bsigma\cdot\bsigma'+g'_0 (\rho)\bsigma\cdot\bsigma'\btau\cdot\btau']
\end{equation}
with
\begin{equation}\label{eq:75}
f(\rho)=f_{ex}+(f_{in}-f_{ex})\rho_0 (r)
\end{equation}
and with the parameters used in most of the previous applications.
\begin{eqnarray}\label{eq:76}\nonumber
f_{in}  =  - 0.002,\; f_{ex} =-1.4,\;
f_{ex}^{\prime}  =  3,0, \;
f_{in}^{\prime}  =  0.76,\\
g  =  0.05,\;
g^{\prime}  =  0.96, \;
C_{0}  =  300\;{\rm MeV fm^{3}}                                
\end{eqnarray}
For the nuclear density $\rho_{0}(r)$ in the interpolation formula we chose the theoretical ground state density distribution
of the corresponding nucleus, 
\begin{equation}\label{eq:77}
\rho_{0}(r) = \sum_{\epsilon_{i} \leq \epsilon_{F}} \frac{1}{4\pi}
(2j_{i} + 1)   R^{2}_{i} (r)
\end{equation}
which is more consistent than the previously used Woods-Saxon distribution. 
For that reason one had to readjust $f_{ex}$ and $f_{ex}^{\prime}$
Here  $R_{i}(r)$ are the single-particle radial wave functions of the  single-particle model used.

For magnetic properties it has been shown \cite{Brown80,Krew88} that one has also to consider the effects of the one-pion exchange.
For that reason on must replace in the spin-isospin channel the conventional Landau-Migdal parametrization by the $\emph{J\"ulich-Stony Brook ansatz}$ which has the following form in the momentum space:
\begin{equation}\label{eq:78}
G'(q)=\int{\frac{d^3k}{(2\pi)^3}}\left[V_{\pi}(\bi{k})+V_{\rho}(\bi{k})\right]\Omega(\bi{q-k})+\delta{G_{0}^{\prime}}
\bsigma\cdot\bsigma'\btau\cdot\btau'
\end{equation}
with
\begin{equation}\label{eq:79}
\Omega(\bi{q})=(2\pi)^3\delta (\bi{q})-\frac{2\pi^2}{q^2}\delta (\left|\bi{q}\right|-q_c)
\end{equation}
Here $V_\pi$ and $V_\rho$ denote the one-pion and one-rho exchange, $\Omega(\textbf{q})$ a parametrization of the nuclear correlation function \cite{Huber69} with $q_c=3.93$, which is the inverse of the Compton wavelength of the $\omega$ meson, and $\delta{G_{0}^{\prime}}$ a small (universal) correction parameter that must be adjusted to the data. 
The $\rho$ exchange has been added because it introduces a natural cut-off for the tensor part of the one-pion exchange. 
A detailed discussion can be found in \cite{Brown80,Krew88}.

The renormalized ph-interaction and therefore also the Landau-Migdal parameters are defined microscopically and can be traced back to the bare nucleon-nucleon interaction. In Ref.\cite{Krew88} these parameters have been calculated from a G-matrix based on a meson exchange model for the bare nucleon nucleon interaction. This is the lowest order for the renormalized ph-interaction
nevertheless the theoretical results were in fair agreement with the phenomenological parameters. The Skyrme ansatz which is mostly used in self consistent nuclear structure calculations has the same form as the original Landau-Migdal parametrization. For that reason, the parameters of the Skyrme ansatz can directly be connected with the Landau-Migdal parameters \cite{BJS75}. These authors have shown that several parameterizations of the Skyrme forces led to instabilities against spin collapse. These relations  can also be used to design new parameterizations for self consistent nuclear structure calculations \cite{Krew77}.
For magnetic properties were the one-pion exchange has to be considered explicitly self consistent calculations based on chiral Lagrangian seems to be more appropriate.

\begin{figure}[htbp]
\begin{center}
\includegraphics[width=7cm]{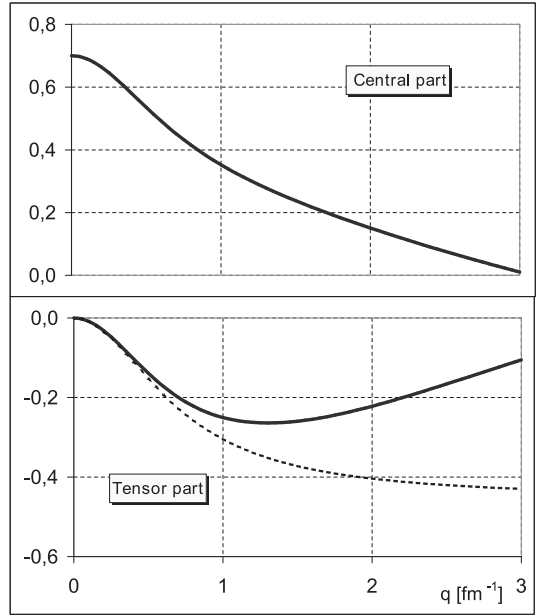}
\end{center}
\caption{\label{fig:2}Momentum dependence of the spin-isospin parameter ${G(q)}'$ in units of $\left[302 MeV fm^3\right]$. The full line is the complete model (\ref{eq:78}), the dashed line in the lower part is the correlated one-pion exchange only.}
\end{figure}

In the upper part of Fig.(\ref{fig:2}) the q-dependence of $G'$ is plotted. 
One realizes that at small momentum transfers the central part of the spin-isospin interaction is strongly repulsive and for larger momentum transfers it is small. This is the reason why, for example, the Gamow-Teller resonances, where small momenta are involved, are strongly shifted to higher energies whereas high-spin states are not \cite{Brown80,Krew88}. 
For the central part $\pi$ -meson and $\rho$ -meson contributions have the same sign and for the tensor part the opposite sign.  
The $\rho$ -meson exchange therefore acts as a natural cut-off for the strong tensor component of the one-pion exchange. 

We have already mentioned that the Migdal parameters have been calculated from a G-matrix starting from the bare nucleon nucleon interaction \cite{Krew88}. In this calculation the authors not only determined the zero-range parameters shows in eq.(\ref{eq:76}) but also the momentum dependence of the spin-isospin interaction given in Fig.(\ref{fig:2}). 

In the TFFS equations effective operators $Q^{\rm eff}$ appear. 
In the case of electric multipole operators one can show using Ward identities that these operators are not renormalized due to the local gauge invariance, as shown in Appendix A.
Magnetic operators are renormalized for two reasons: 
(I) there exists no conservation law for the spin contribution and 
(II) there are mesonic contributions. 
For that reason one has to parametrize the magnetic operators \cite{Rev77}, the parameters of which are universal for all medium and heavy mass nuclei.
\begin{figure}[htbp]
\begin{center}
\includegraphics[width=7cm]{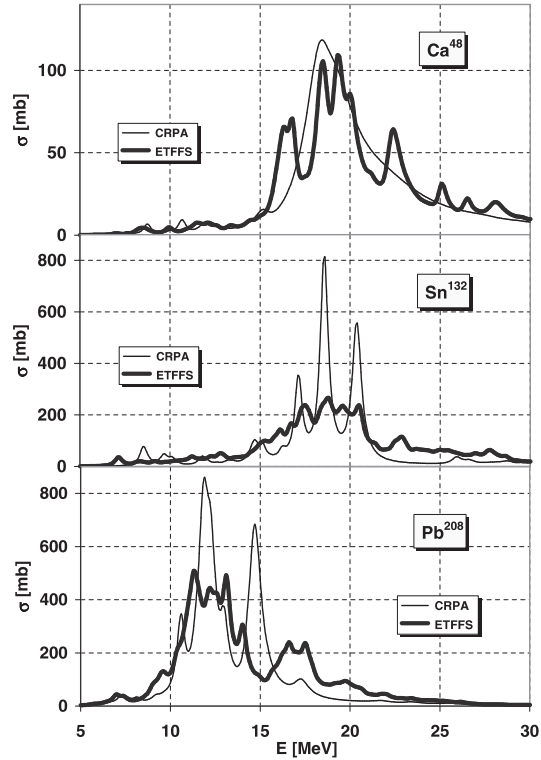}
\end{center}
\caption{\label{fig:3}E1 photo absorption cross section for three double magic nuclei. The thin lines denote the results of the continuum RPA and the thick lines the results of the extended theory.}
\end{figure}
In Fig.(\ref{fig:3}) the results of TFFS calculations for the electric giant dipole resonances in $^{40}$Ca, $^{132}$Sn and $^{208}$Pb are shown. 
The single particle continuum is included in an exact way. 
In order to simulate the experimental resolution and more complex configurations, a smearing parameter $\Delta$ =250 keV is used. 
The (isovector) electric dipole operator is given as: 
\begin{eqnarray}\label{eq:80}
Q_{1M}=\frac{eN}{A}\sum_{i=1}^{Z}r_{i}Y_{1M}(\Omega_i)-\frac{eZ}{A}\sum_{i=1}^{N}r_{i}Y_{1M}(\Omega_i)
\end{eqnarray}  
In $^{208}$Pb, where the electric giant dipole resonance is known \cite{HvW01}, the peak energies agrees with the experimental centroid-energies (up to 1 MeV) and the resonance exhaust a large fraction of the energy weighted sum rule.
However, the theoretical width is much too small and the maximum of the cross section is much too high because the spreading width, which is due to the fragmentation of the single particle strength, is not considered. The same is true for the other two cases if one compares the theoretical results with neighboring nuclei data exist.
In the next section we will discuss an extension of the TFFS, where the low lying collective phonons are coupled to the single particle states, which gives rise to a fragmentation of the single particle strength. 
In this way we include, as we will show, a major part of the spreading width.

\section{Extended theory of finite Fermi systems (ETFFS)}
As mentioned earlier, the original TFFS allows one to calculate only the centroid energies and total transition strength of giant resonances because the approach is restricted to 1p1h configurations. 
In order to describe nuclear structure properties in more detail one has to include 2p2h, or even more complex configurations. A theoretical approach that takes into account the complete 2p2h configuration space and a realistic 2p2h interaction is numerically intractable if one uses a realistically large configuration space.
For this reason the main approximation in the ETFFS concerns the selection of the most important 2p2h configurations. 
One knows from experiments, e.g. from the neighboring odd mass nuclei of $^{208}$Pb, that the coupling of low-lying collective phonons to the single particle states gives rise to a fragmentation of the single-particle strength, which is seen in  even-even nuclei as a spreading width in the giant resonances.
In the past 15 years an extension of the TFFS that includes, in a consistent microscopic way, the most collective low-lying phonons has been performed by Tselyaev \cite{Ts89} and Kamerdzhiev, Speth and Tertychny \cite{Rev04} (and references therein). 
In this way one considers a special class of 2p2h configurations because the phonons are calculated within the conventional (1p1h) TFFS. 
The phonons give rise to a modification of the particle and hole propagators, the ph-interaction and the ground state correlations. 
The extension that includes uncorrelated 2p2h configurations is discussed in ref. \cite{Droz90} and \cite{Kam97}.

\subsection{Some basic relations}
In Landau's theory the self energy $\Sigma$ is irreducible in the one-particle and one-hole channels, respectively, and the kernel $K$ in the integral equation for the response function (see Fig.(\ref{fig:1})) is irreducible in the particle-hole channel.
One introduces now a hierarchy of energy dependencies: 
As one can see from eq.(\ref{eq:53}) the particle-hole propagator introduces a strong dependence in the energy transfer $\Omega$. 
Compared to this singular behavior one neglects in Landau's theory the energy dependence in the ph-interaction. 
One neglects consistently the energy dependence of the self-energy $\Sigma$ in the Dyson equation (eq.(\ref{eq:33}) and considers the quasi-particle and quasi-hole poles only.
This approach is of leading order in the energy transfer $\Omega$. 
In the extended theory, where phonons are introduced, one considers the next-to-leading order in the energy transfer; i.e., the self energy and the ph-interaction become energy dependent.
We the write, explicitly,
\begin{equation}\label{eq:81}
\widetilde{\Sigma}_{\nu}(\epsilon)= \Sigma_{\nu} +{\Sigma}^{ext}_{\nu}(\epsilon)
\end{equation}
\begin{equation}\label{eq:82}
\widetilde{F}^{ph}_{\nu_1 \nu_3,\nu_2 \nu_4}(\Omega)=F^{ph}_{\nu_1 \nu_3,\nu_2 \nu_4}+F^{ph,ext}_{\nu_1 \nu_3,\nu_2 \nu_4}(\Omega)
\end{equation}
\begin{figure}[htbp]
\begin{center}
\includegraphics[bb=44 463 795 551,width=12cm]{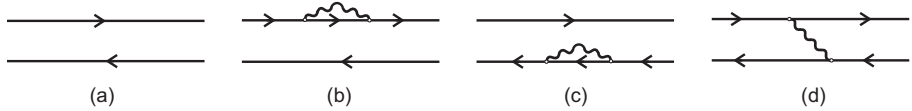}
\end{center}
\caption{\label{fig:4}Graphs corresponding to the propagator of the (a) RPA and (b-d) the extended theory. The wavy lines and the thin lines denote the phonons and the single particle propagators, respectively.}
\end{figure}
First we discuss the so-called $g^2$ approximation \cite{Rev04,Rev77}, which is graphically shown in Fig.(\ref{fig:4}). 
The upper part of  graph (b) gives a correction to the Dyson equation for quasi-particles, which now has the form
\begin{equation}\label{eq:83}
\left[\epsilon -{\epsilon_{\nu_1}}-\sum_{\nu_2 ;i}
\frac{\left| \gamma^{\nu_2 ;i}_{\nu_1}\right|^2}
{\epsilon -\Omega_i  -\epsilon_{\nu_2}} \right]g_{\nu_1}^{ext}(\epsilon )=1
\end{equation}
Here $g_{\nu}^{ext}(\epsilon )$ is the one-quasi-particle Green function in the extended theory and the vertex $\gamma_{\nu} ^{\mu;i}$, which couples the quasi-particle state $\nu$ to the core excited configuration $\mu{\otimes}i$, is given by
\begin{equation}\label{eq:84}
\gamma_{\nu} ^{\mu;i} = \sum_{\alpha,\beta} F^{ph}_{\nu \alpha; \mu \beta} \chi^{i}_{\alpha\beta}
\end{equation}
where $\chi^i$ is the RPA wave function of the phonon considered.
The corresponding energy-dependent correction for the ph-interaction has the form
\begin{equation}\label{eq:85}
F^{ph,ext}_{\alpha \mu, \beta \nu}(\Omega,\epsilon,\epsilon') \nonumber
= \sum_{i}\frac{(\gamma^{\mu;i}_{\alpha})^* \gamma^{\nu;i}_{\beta}}{ \epsilon - \epsilon'+  (\Omega_i-i\delta)}
\end{equation}
This model has been applied, for example, to M1-resonances in closed shell nuclei \cite{Kam89}. 
This approximation has, in general, problems with the so-called second order poles, which give rise to a distortion of the strength function near these poles.  
For this reason the model has been extended in several steps: 
(I) summation of the $g^2$ terms \cite{Kam86} and
(II) partial summation of diagrams \cite{Ts89}, which is call $\emph{chronological decoupling of diagrams}$. 
In the latter case all $1p1h\otimes phonon$ contributions are consistently included and all more complex configurations are excluded --as long as one neglects ground-state correlations. The actual formulas, however, include the ground state correlations
completely.
In the present approach, like in \cite{Ts89} two types of ground state correlations are included:
the conventional RPA ground state correlations, which affect the location and the magnitude of the residua of the ph propagator only, and the new type of ground state correlations, which is caused by the phonons. 
These correlations are qualitatively different from the conventional RPA correlations because they create new poles in the propagator, which then cause transitions between the $1p1h\otimes phonon$ in the ground state and the excited states. 
They give rise to a qualitative change of the strength distribution and a change in the sum rules for the moments of the strength function.  
As in the TFFS, the energy dependence of the generalized propagator $A^{ext}$ is much stronger than that of the generalized ph interaction, as the next-to-leading order energy dependence is removed from the interaction and explicitly taken into account in the generalized propagator. Therefore one considers explicitly only the energy dependence of the propagator and neglects the energy dependence of the ph interaction. The interaction is parametrized as before; the corresponding parameters may differ from the previous ones.
The final equation for the change of the quasi-particle density in an external field has the identical form to that in the TFFS (eq.(\ref{eq:66})
\begin{eqnarray}\label{eq:86}
\widetilde{\rho}^{ext}(\bi{r},\Omega) =&-\int d^{3}\bi{r}'A^{ext}(\bi{r},\bi{r}',\Omega) Q^{eff}(\bi{r}',\Omega)d^{3}\\ \nonumber
&-\int d^3\bi{r}'d^3\bi{r}''A^{ext}(\bi{r},\bi{r}',\Omega) F^{ph,}(\bi{r}',\bi{r}'')\widetilde{\rho}^{ext}(\bi{r}'',\Omega) 
\end{eqnarray}
The analytical form of the generalized propagator can be found in Ref.\cite{Rev04}.

\subsection{Numerical applications}
As in the case of the TFFS the numerical input to the ETFFS are two sets of phenomenological parameters that describe 
(I) the Woods-Saxon single particle potential and 
(II) the effective ph interaction. 
If one includes phonons, the problem of double counting appears in the two input data sets. 
One therefore has to subtract explicitly the phonon contribution from the mean field solutions as well as the ph interaction. In order to obtain the new \emph{refined} single particle basis $\left\{\widetilde{\epsilon}_\lambda,\widetilde{\phi}_\lambda\right\}$ one has to solve the equation
\begin{equation}\label{eq:87}
\widetilde{\epsilon}_{\nu_1} =\epsilon_{\nu_1} -\sum_{\nu_2 ;i}
 \frac{\left| \gamma^{\nu_2 ;i}_{\nu_1}\right|^2}
{\epsilon_{\nu_1} -\Omega_i  -\widetilde{\epsilon}_{\nu_2}}
\end{equation}
where the second term on the right side is graphically given by the upper part of graph (b) in Fig.(\ref{fig:4}). 
As the wave functions are only little changed, we restrict the refinement to the single particle energies. 
In Fig.(\ref{fig:3}) we also show the ETFFS-results  for $^{48}$Ca, $^{132}$Sn and $^{208}$Pb.
The effect of the phonon coupling is very clearly demonstrated in all three cases. 
The phonon coupling reduces the maximum of the strength distribution and shifts part of the strength to higher energies. 
This effect increases the width of the resonances and creates the long tails, which are seen in the data. For $^{208}Pb$ the theoretical mean energy has the value $\widetilde{E_{th}}= 12,84 MeV$ compared with experimental value $\widetilde{E_{exp}} = 13.4 MeV$ and the theoretical width $\Gamma_{th}= 3,6 MeV$  compared to $\Gamma_{exp}=4,1 MeV$. For 
The energy weighted sum rule (EWSR) is only little changed in the range between 5-30 MeV. For $^{48}$Ca holds: $\widetilde{E_{th}}= 19,71MeV$ compared to $\widetilde{E_{exp}}= 19,6 MeV$ and $\Gamma_{th}= 5,36 MeV$  compared to $\Gamma_{exp}=7,1 MeV$. The phonon coupling improves the CRPA results by a factor of two.
\subsection{A realistic example for the application in data analysis}
The renewed interest in nuclear structure physics comes from newly built and planned experimental facilities that will give new information on nuclei far from stability.
One of the main reasons for this is its application to astrophysics.
Here, for example, one would like to know the compression modulus  and the symmetry energy of neutron rich nuclei. 
Such experiments will be performed in so-called inverse kinematics; i.e. the radioactive isotopes will be accelerated and will be scattered from proton, deuteron or alpha particle targets.
The analysis of such experiments will be very complicated as the breathing mode and giant electric dipole resonance are expected to be broad and because various multipole resonances overlap each other.
For that reason on needs realistic strength distributions and energy dependent microscopic transition densities as input into the scattering codes in order to be able to extract reliable physical results.
In the following we demonstrate this in the analysis of $(\alpha,\alpha')$ cross sections in $^{58}$Ni, where the isoscalar breathing mode has been investigated.
\begin{figure}[htbp]
\begin{center}
\includegraphics[width=9cm]{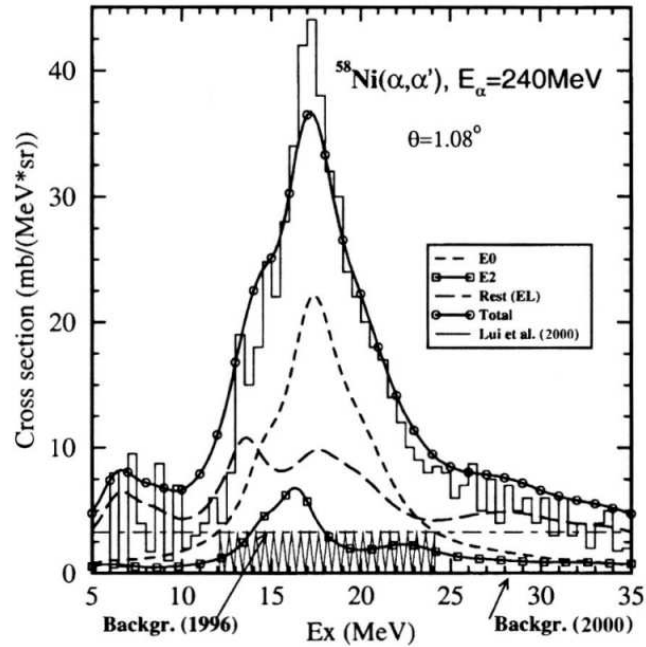}
\end{center}
\caption{\label{fig:5}Cross sections of $^{58}$Ni$(\alpha,\alpha')$ at $1.08^\circ$. At this angle the cross section is dominated by the isoscalar monopole resonance.}
\end{figure}
A general problem in nuclei with $A<$ 90 is that the isoscalar monopole resonance is very broad and no longer concentrated in one single peak.
In deformed nuclei (such as one expects for nuclei far from stability) the giant resonances are split due the deformation, which gives rise to additional complications \cite{Zaw78}.
In lighter nuclei the role of the surface becomes more important than in the heavier ones and, in addition, $(\alpha,\alpha')$ experiments are most sensitive to the nuclear surface.
For these reasons the correct treatment of the surface in those nuclei is crucial.
All this has to be considered if one develops models for giant resonances in light and medium mass nuclei for which the conventional RPA approach is extended to include surface modes and the single-particle continuum.
Compared to this extension of the configuration space, the way that one determines the mean field and the residual interactions is of less importance.
When the present example was first published the experimental situation in $^{58}$Ni was rather unclear.
In the original analysis of the $(\alpha,\alpha')$ data $\cite{You96}$ only 32\% of the EWSR was observed.
The authors used standard data analysis techniques with the same phenomenological transition densities for the whole energy range.
For comparison the same type of experiment in $^{40}$ Ca and a similar analysis by the same authors showed $(92\pm15)$ \% of the EWSR \cite{You97}.
With an improved --but still conventional-- analysis the authors of Ref.\cite{Sat97} obtained about 50 percent.
Such differences might have serious consequences for nuclear matter compressibility and its application to astrophysics.
In what follows we discuss an analysis of the $(\alpha,\alpha')$ experiments in $^{58}$Ni, where the isoscalar giant resonance has been analyzed using the results of ETFFS calculations.
Within this model the authors calculated the distribution of the transition strengths of the isoscalar E0-E4 resonances and the corresponding energy dependent transition densities. Using these transition densities one obtains, in the standard way, $(\alpha,\alpha')$ scattering cross sections that were compared with the data in the energy range between 5-35 MeV \cite{Lui00}.
One of the problems that occurs in such analysis is the subtraction of the \emph{instrumental} background.
Usually the experimentalists draw a straight line between the beginning and end of the resonance peak, where the strength distributions become flat. It has been demonstrated in Ref.\cite{Kam00} that if one adds up the contributions of the various \emph{weak} multipole contributions one also obtains a flat back ground, which may erroneously be interpreted as part of the \emph{instrumental} background.
\begin{figure}[htbp]
\begin{center}
\includegraphics[width=9cm]{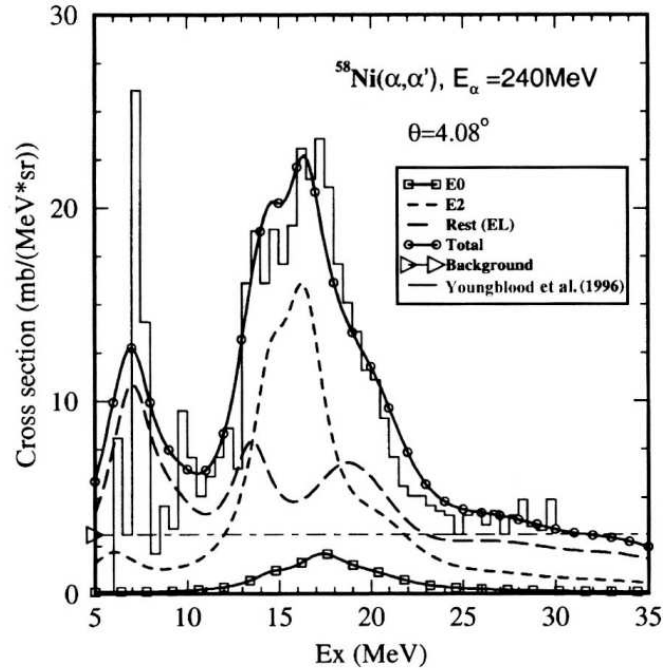}
\end{center}
\caption{\label{fig:6}Cross sections of $^{58}$Ni$(\alpha,\alpha')$ at $ 4.08^\circ$. At this angle the cross section is dominated by the isoscalar quadrupole resonance}
\end{figure}
This \emph{background} includes also a fraction of the \emph{resonance} strength that one wishes to extract.
In Fig.(\ref{fig:5}) the differential $(\alpha,\alpha')$ cross section for the scattering angle $\Theta =1.08^{\circ}$ is shown, where the E0 contribution dominates the cross section. The full theoretical result (full line with dotes)\cite{Kam00} is compared with the experimental cross section \cite{Lui00}. These data where obtained by subtracting an \emph{instrumental} back ground denoted by "Backgr(2000)" from the original experimental spectrum. The theoretical cross section is the sum of five different multipole resonances, of which the E0 and E2 contributions are shown explicitly. The straight line in the lower part of the figure, denoted by " Backgr.(1996)" corresponds to the analysis of Ref.\cite{You96}, where the energy interval was smaller (12-25 MeV). In this analysis only 32\% of the EWSR E0 was found, whereas the present theory predicts 71.4\% of the E0 EWSR in this regime. If one considers the larger 12-35 MeV interval, the theoretical strength exhaust 89.6\%.
We also show data taken at $\Theta =4.08^{\circ}$, where the E0 resonance has a minimum and the E2 its maximum. 
The same analysis as before was performed by Ref.\cite{Kam00} and the result is shown in Fig.(\ref{fig:6}).
Here the E2 strength dominates the cross section whereas the breathing mode contributes only a tiny fraction.  The \emph{other} isoscalar modes again give an important contribution to the cross section. As both experimental cross sections are well-reproduced within the ETFFS analysis one may conclude that the major part of the E0 and E2 EWSR strength is in the energy range considered. One may also conclude that realistic microscopic nuclear structure models are necessary for the analysis of experiments in the giant resonance region. The models have to go beyond 1p1h RPA and must include the coupling to low-lying phonons in order to reproduce the width of giant resonances in a realistic manner.

\section{Conclusion}
We have reviewed Migdal's \emph{Theory of Finite Fermi Systems} (TFFS) and the \emph{Extended Theory of Finite Fermi Systems} (ETFFS) as well as their application to nuclear structure. 
Both approaches are derived within the Green function formalism, which enables one to obtain the final equations without  approximations. 
For that reason one can apply conservations laws, which restrict the number of parameters in the actual numerical application. 
The TFFS and the ETFFS are frameworks that require single-particle energies,single-particle wave functions and the ph interactions as input. 
The single particle basis may be obtained from phenomenological single-particle models, as discuss here, or from self consistent mean field calculations. 
The  original Landau Migdal ph-interaction includes delta functions and derivatives of delta functions in $\textbf{r}$ -space, so that Skyrme forces can be directly connected with the Landau Migdal interaction  \cite{BJS75} and \cite{Krew77}. 
This fact can be used as a test for the various versions of the Skyrme parameters but also for designing new set of parameters which are especially appropriate for self consistent nuclear structure calculations. 
The original Landau Migdal theory is formally identical with the  1p-1h RPA. 
Therefore if one calculates in this approach giant multipole resonances one  can only reproduce the mean energy and the total strength of those states. 
For a quantitative description an extension of the theory is necessary. 
One has to include in a consistent way the effects of the phonons. 
This has been done in the \emph{Extended Theory of Finite Fermi Systems} which we also reviewed here. 
We have pointed out that if one uses experimental single-particle energies and/or a correspondingly adjusted single particle model in the ETFFS, one has to correct for the phonon contributions in the single particle spectrum as the phonons (or, in general, 2p2h configurations) give rise to a compression of the single particle spectrum \cite{Droz90}. 
Therefore a self consistent mean field approach provides the correct input data because in this case the phonon contributions are not included. 
For the same reason is the conventional RPA not the appropriate approximation in  self consistent calculations because the uncorrelated ph spectrum is in general too wide as a consequence of the phonon contribution being missing. 
In the case of very collective resonances, which are strongly shifted in energy, this deficiency can be removed by an appropriate ph interaction. W
e are aware of the fact that, e.g. for a specific Skyrme parametrization the residual interaction is defined by the Skyrme interaction itself and one has, in principle, no freedom. 
But we are also aware of the fact that there exist dozens of different parameter sets that give similar ground state properties but quite different nuclear structure results \cite{Krew77}. 
Moreover, the size of the configuration space can also be chosen in an appropriate way. 
Thus the arbitrariness in self-consistent calculations lies in the choice of the effective Lagrangian, effective Hamiltonian or energy-density functional and the size of the configuration space. 
For less collective states, such as the pygmy resonances, the energies of which are only little shifted by the residual interaction, self consistent calculations in general have difficulties reproducing simultaneously these experimental data and the giant electric dipole resonance. 
Therefore some authors, e.g.\cite{Len04}, $\emph{compress}$ the too large level spacing by multiplying it \emph{adhoc} with the corresponding effective mass in order to fit the data. 
This may be justified by the energy dependence of the effective mass but it certainly destroys the self consistent treatment. We have also mentioned that for magnetic properties one has to consider in the spin-isospin channel the one pion exchange explicitly. 
For that reason self consistent calculations, starting from chiral Lagrangian may be the more consistent approach compared to the Skyrme Hamiltonian.
 
Acknowledgments: 
We thank John Durso, Ulf Mei{\ss}ner and Victor Tselyaev for carefully reading the manuscript and many useful comments. 
We also appreciate the continuous collaboration with Sergej Kamerdzhiev and Gennady Tertychny in this field.
The work was supported in part by the DFG grant No. GZ:436RUS113/806/0-1.
  
\section{Appendix A: Ward identities}
As the final equations of the Landau-Migdal theory are derived without approximations, one can apply conservations laws in order to restrict the number of free parameters. 
Here we show, using local gauge invariance, that scalar vertices are not renormalized, i.e. one has to use the bare charges (for protons and neutrons) for the electric multipole operators. 
We start with the one-particle Green function:
\begin{equation}\label{eq:88}
g(x,y) = -i<A0|T \left\{\bPsi\left(x\right)\bPsi^{+} \left(y\right)\right\}|A0>
\end{equation}
where x and y denote four-vectors. The gauge transformation of the $\Psi$-operators where $q$ is the charge
\begin{equation}\label{eq:89}
\bPsi'(x) = e^{iq f(x)} \bPsi(x)\qquad   \bPsi'^+ (x) = e^{-iq f(x)} \bPsi^+ (x)
\end{equation}
leads to the following change in the Hamiltonian 
\begin{equation}\label{eq:90}
\delta H(x) = - j^{i}(x)  \frac{\partial f(x)}{\partial x_i}
\end{equation}
Here $j_{n,p}^i = (j_{n,p}^0;j_{n,p}^\alpha )$ is the four-current of neutrons and protons. 
This field does not give rise to any physical changes.
If one expands the transformed operators in the one-particle Green function up to first order, one obtains 
\begin{equation}\label{eq:91}
\delta g(x,y) = i(q f(x)g(x,y)-g(x,y)q f(y))
\end{equation}
The inverse of this equation has the form
\begin{equation}\label{eq:92}
\delta g^{-1} =-g^{-1}\delta g g^{-1} =-i(g^{-1}(x,y)q f(y)-q f(x)g^{-1}(x,y))
\end{equation}
On the other hand one can calculate the change of the Green function due to the additional term in the Hamiltonian eq.(\ref{eq:90}) with help of the linear response function
\begin{equation}\label{eq:93}
\delta g(x,y) = \int dz \: L(x,z;y,z) \delta H(z)
\end{equation}
The corresponding inverse change of the Green function is given as:
\begin{equation}\label{eq:94}
\delta g^{-1}(x,y) = \int dz \:\tau (x,z;y,z)j^{i}(z) \frac{\partial f(z)}{\partial z_i}\,
\end{equation}
With the field: 
\begin{equation}\label{eq:96a}
f(z)= e^{i\left(\bi{kz}-\Omega t\right)}
\end{equation}
one obtains from eq.(\ref{eq:92}) and eq.(\ref{eq:94})
\begin{eqnarray}\label{eq:96}
-i(g^{-1}(\bi{x},\bi{y}; \omega +\frac{\Omega}{2})q e^{i\bi{k}\bi{y}}-q e^{i\bi{k}\bi{x}}g^{-1}(\bi{x},\bi{y}; \omega -\frac{\Omega}{2})) =\\ \nonumber
-i\Omega\int \!d^{3}z \; \tau(\bi x,\bi z ;\bi y,\bi z;\omega,\Omega )\rho(\bi z)e^{i\bi k\bi z}-i\int \!d^3 z \; \tau(\bi x,\bi z ;\bi y,\bi z;\omega,\Omega )\left(\bi j(\bi z)\bi k\right) e^{i\bi k\bi z}
\end{eqnarray} 
This equation has in the single particle basis $\varphi _{\nu}(\bi x)$ the form:
\begin{eqnarray} \label{eq:97} \fl
\sum _{\nu_3} \left\{\left\langle {\nu _1} \right| g^{-1}(\omega +\frac{\Omega}{2})         \left|{\nu _3} \right\rangle{\left\langle{\nu _3} \right|{q  e^{i \bi k \bi r}} \left|{\nu _2}\right\rangle } -{\left\langle{\nu _1} \right|{q  e^{i \bi k \bi r}} \left|{\nu _3}\right\rangle }\left\langle {\nu _3} \right| g^{-1}(\omega -\frac{\Omega}{2})         \left|{\nu _2} \right\rangle\right\} \\ \nonumber
= \Omega \left\langle {\nu _1}\right|\widetilde{\tau} (\omega, \Omega ; \left[ { q e^{i \bi k \bi r}}\right])\left|{\nu _2} \right\rangle +\left\langle {\nu _1}\right|\widetilde{\tau} (\omega, \Omega ; \left[ (\bi jk){ e^{i \bi k \bi r}}\right])\left|{\nu _2} \right\rangle
\end{eqnarray}
with
\begin{equation}\label{eq:98}
\left\langle {\nu _1}\right|\widetilde{\tau} (\omega, \Omega ; \left[ { q e^{i \bi k \bi r}}\right])\left|{\nu _2} \right\rangle = \sum _{\nu _3 \nu _4} \widetilde{\tau} _{\nu _1 \nu _3 \nu _2 \nu _4}(\omega , \Omega )\left[q e^{i\left(\bi{kr}\right)}\right]_{\nu_4\nu_3}
\end{equation}
and
\begin{equation}\label{eq:99}
\left\langle {\nu _1}\right|\widetilde{\tau} (\omega, \Omega ; \left[(\bi jk) { e^{i \bi k \bi r}}\right])\left|{\nu _2} \right\rangle = \sum _{\nu _3 \nu _4} \widetilde{\tau} _{\nu _1 \nu _3 \nu _2 \nu _4}(\omega , \Omega )\left[(\bi jk) e^{i\left(\bi{kr}\right)}\right]_{\nu_4\nu_3}
\end{equation}
We have replaced in eq.(\ref{eq:97}) the charge distribution $\rho$ by the point charge $q$.
With the singular part of the one particle Green function eq.(\ref{eq:53})
one obtains a relation between the bare operators and the vertices $\widetilde{\tau }$
\begin{eqnarray} \label{eq:100}
\left\{\frac{\omega+\frac{\Omega}{2}-\epsilon _{\nu _1}}{z_{\nu_1}} -\frac{\omega-\frac{\Omega}{2}-\epsilon _{\nu _2}}{z_{\nu_2}}\right\}\left\langle \nu _1 \right| q e^{i\bi kr}\left| \nu _2 \right\rangle =  \\ \nonumber \\ \nonumber
\Omega \widetilde{\tau} _{\nu _1 \nu _2}\left[q e^{i\bi kr}\right]+\:\widetilde{\tau} _{\nu  _1 \nu _2}\left[(\bi j \bi k)e^{i\bi kr }\right] \nonumber
\end{eqnarray}
We multiply this equation with $\sqrt{z_{\nu _1  \nu _2}}$ and replace the vertices $\widetilde{\tau}$ by the renormalized vertices $\tau$ defined in eq.(\ref{eq:58})
\begin{equation}\label{eq:101}  \fl
(\Omega-\epsilon_{\nu_1}-\epsilon_{\nu_2})\left\langle \nu _1\right|(q e^{i\bi kr}\left|\nu_2 \right\rangle = \Omega \; \tau _{\nu_1 \nu_2}\left[q e^{i \bi kr}\right] + \tau _{\nu_ 1 \nu_2} \left[ \bi jk e^{i \bi kr}\right]
\end{equation}
Here the approximation $\sqrt{\frac{z_{\nu _1}}{z_{\nu_2}}}\approx\sqrt{\frac{z_{\nu _2}}{z_{\nu_1}}}$
was used. 
If one puts the renormalized vertex function (eq.(\ref{eq:102})) explicitly into eq.(\ref{eq:101})
\begin{eqnarray}\label{eq:102}
\tau_{\nu_1\nu_3, \nu_2 \nu_4 }\left( \omega ,\Omega\right)=\tau^{\omega}_{\nu_1 \nu_3, \nu_2 \nu_4 }\left( \omega , \Omega\right) +\\ \nonumber
\sum_{\nu_5 \nu_6}F^{ph}_{\nu_1 \nu_5 , \nu_2 \nu_6}\left( \omega,\frac{\epsilon_{\nu_5}+ \epsilon_ {\nu_6}}{2}, \Omega\right) \; \frac{n_{\nu_5} -n_{\nu_6}}{\epsilon_{\nu_5} - \epsilon_{ \nu_6} -\Omega } \;\tau_{\nu_5 \nu_3, \nu_6 \nu_4 }\left( \frac{\epsilon_ {\nu_5} + \epsilon_{\nu_6}}{2} ;\Omega\right)
\end{eqnarray}
and considers the limit $\Omega >> (\epsilon _{\nu_1}-\epsilon_{\nu _2})$. One obtains the final result: 
\begin{equation}\label{eq:103}
\sum_{\nu_3 \nu_4}\tau^{\omega}_{\nu_1 \nu_3
\nu_2\nu_4}(q e^{i\left(\bi{kr}\right)})_{
\nu_3\nu_4}=(q e^{i\left(\bi{kr}\right)})_{\nu_3\nu_4}\,.
\end{equation}
Scalar (electric multipole) operators are not renormalized.

\end{document}